\documentclass[twoside,12pt]{article}
\usepackage{epsfig}

\def\Journal#1#2#3#4{{#1} {#2} (#4) #3 }

\def\NPA{{\em Nucl. Phys.} A}

\def\PLB{{\em Phys. Lett.} B}

\def\PRL{\em Phys. Rev. Lett.}

\def\PREP{\em Phys. Rep.}

\def\PRC{{\em Phys. Rev.} C}

\def\ZPA{{\em Z. Phys.} A}

\def\APP{{\em Acta Physica Polonica} B}

\newcommand{\be}{\begin{equation}}
\newcommand{\ee}{\end{equation}}
\newcommand{\bea}{\begin{eqnarray}}
\newcommand{\eea}{\end{eqnarray}}

\begin{document}

\title{ \vspace{1cm} Production of fragments with and without strangeness within a 
combined BUU+SMM approach}
\author{T.\ Gaitanos, H.\ Lenske, U.\ Mosel
\\
Instit\"ut f\"ur theoretische Physik, Universit\"at Giessen, Giessen, Germany}
\maketitle
\begin{abstract} 
The formation of hypernuclei in hadron-induced reactions and 
in heavy-ion collisions within a combination of a covariant transport 
model and a statistical fragmentation approach is investigated. We 
study the applicability and limitations of such a hybrid approach by 
comparing fragmentation data in heavy-ion collisions and proton induced 
reactions. It turns out that the combined approach describes fairly well 
multiplicities and momentum spectra of fragments. We thus extend the 
model by including strangeness degrees of freedom in the fragmentation 
process, modeled by a phase-space coalescence method. We provide theoretical 
predictions on spectra and on inclusive cross sections of light hypernuclei 
for the future experiments on hypernuclear physics at the new GSI facility.
\end{abstract}

\section{Introduction}

Hypernuclear physics opens the unique opportunity to study the properties of 
the hyperon-nucleon and hyperon-hyperon interaction \cite{hyp1}. The theoretical 
production of light hypernuclei in reactions was originally proposed by 
Kerman and Weiss \cite{kerw}. Since then this topic has been attracted again 
theoretical interest \cite{wakai1}, motivated by the new FAIR facility 
at GSI, in which projects on hypernuclear physics are running or under planing 
\cite{hypHI,PANDA}.

In this work the initial non-equilibrium stage of a reaction is described by a 
covariant transport theory of Boltzmann type (Giessen-BUU), while the 
fragmentation mechanism of the final channel is modeled by a purely statistical 
approach (Statistical Multifragmentation Model, SMM \cite{SMM}). The first 
stage of a dynamical process is modeled by the BUU transport equation until 
the system approaches an intermediate equilibrated stage, which may be an excited 
configuration. The de-excitation of this configuration is then statistically 
treated by the SMM model, which contains different fragmentation mechanisms. 
We show that the combined approach reproduces fairly well fragment multiplicities 
and spectra in proton induced reactions and in spectator fragmentation in intermediate 
energy heavy-ion collisions. The model is then applied to higher energies by 
allowing the formation of fragments with strangeness degrees of freedom. 
Theoretical predictions on light single hypernuclei for specific reactions 
related to projects of the future FAIR facility are then discussed.

\section{The hybrid GiBUU+SMM model}

The standard theoretical approach in describing collisions induced by hadrons 
or heavy-ions is based on the semiclassical kinetic theory of statistical 
mechanics \cite{kada}. Here we use the covariant analogon of this equation 
known in the literature \cite{RBUU} as the Relativistic 
Boltzmann-Uehling-Uhlenbeck (RBUU) equation. 
\begin{eqnarray}
& & \left[
k^{*\mu} \partial_{\mu}^{x} + \left( k^{*}_{\nu} F^{\mu\nu}
+ M^{*} \partial_{x}^{\mu} M^{*}  \right)
\partial_{\mu}^{k^{*}}
\right] f(x,k^{*}) = {\cal I}_{coll}
\label{rbuu}
\quad .
\end{eqnarray}

The transport equation (\ref{rbuu}) describes the space-time evolution of 
the $1$-body phase space distribution function $f(x,k^{*})$ for the 
different types of hadrons (nucleons including their higher resonant 
excitations, pions, kaons, and other mesons). The numerical implementation 
of the transport equation (\ref{rbuu}), as realized the Giessen-group, is 
called the Giessen-BUU (GiBUU) equation \cite{GiBUU}. Details on the approach 
and the model parameters can be found in \cite{Lario,gait08}. 

\begin{figure}[tb]
\begin{center}
\begin{minipage}[t]{10 cm}
\epsfig{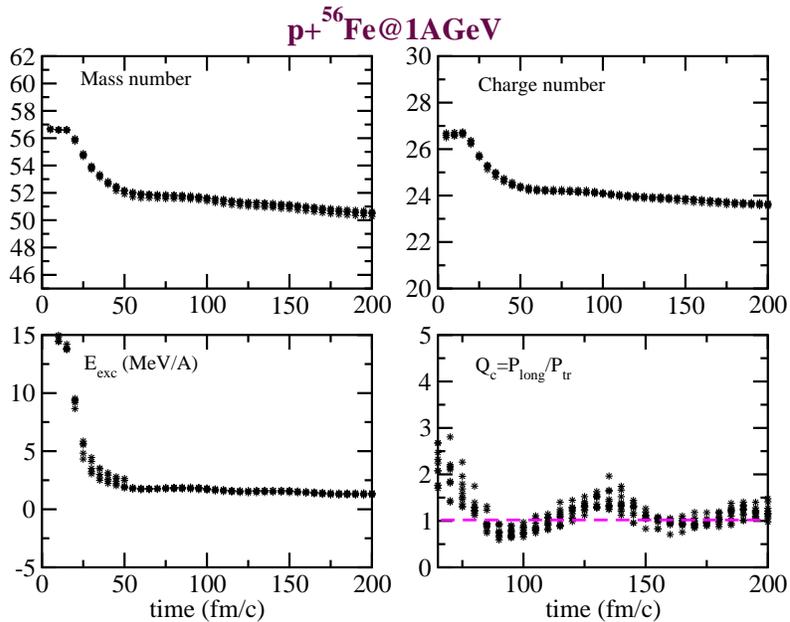}
\end{minipage}
\begin{minipage}[t]{16.5 cm}
\caption{Time evolution of a residual nucleus \cite{residual} in a central 
$p+Fe@1~GeV$ reaction. Displayed are the average mass number (top-left), 
average charge number (top-right), excitation energy (bottom-left) and the 
anisotropy ratio (bottom-right) as function of time. 
\label{Fig1}}
\end{minipage}
\end{center}
\end{figure}
The fragmentation process, which is crucial for the production of hypernuclei, 
is not physically accounted by the transport equation. The standard approach 
in modeling fragmentation within transport studies is the phenomenological 
phase-space coalescence picture \cite{coala}. However, in situations with long 
time scales for the fragmentation process, as spectator fragmentation in 
heavy-ion collisions and dynamics in hadron-induced reactions, a statistical 
description is necessary. Here we apply the SMM model in describing 
the fragmentation of spectator matter in heavy-ion collisions or of residual 
nuclei in hadron-induced reactions. The physical condition of passing from 
the dynamical (BUU) to the statistical picture (SMM) is restricted to the 
existence of a pre-equilibrium excited configuration, determined by the 
anisotropy ratio of the local longitudinal and transverse pressure 
components. This is demonstrated in Fig.~\ref{Fig1} for a proton-induced 
reaction. After the initial non-equilibrium phase, in which the proton beam 
penetrates the nucleus and excites it, the residual system achieves an equilibrated state 
at $t\sim 75~fm/c$, characterized by constant values of mass, charge number 
and excitation energy. Also the anisotropy ratio approaches unity. 
However, the SMM approach does not account for 
strangeness degrees of freedom, thus we apply a phase-space coalescence 
algorithm for the formation of hyperfragments, in which the coalescence 
parameter in coordinate {\it and} in momentum space are fixed to results of 
analytical studies by Wakai et al. \cite{wakai1}. 

\section{Fragmentation at low energies}

It is convenient to study first the reliability of the combined approach 
by means of fragmentation observables at low incident energies, 
before applying it to the production of hypernuclei. The main quantities 
for the formation of hypernuclei are the cross sections of light fragment 
production and those of strangeness production. 

\begin{figure}[tb]
\begin{center}
\begin{minipage}[t]{10 cm}
\epsfig{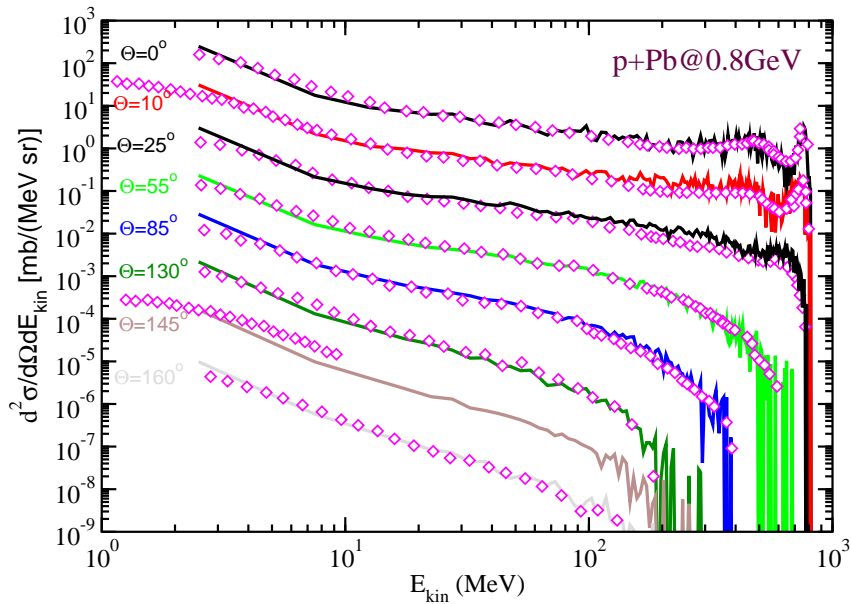}
\end{minipage}
\begin{minipage}[t]{16.5 cm}
\caption{Double differential cross sections for emitted neutrons in 
$p+Pb@0.8~GeV$ reactions. Theoretical calculations (solid lines) are compared 
with experimental data (open diamonds) taken from \protect\cite{spectraEXP}.
\label{Fig2}}
\end{minipage}
\end{center}
\end{figure}
\begin{figure}[tb]
\begin{center}
\begin{minipage}[t]{10 cm}
\epsfig{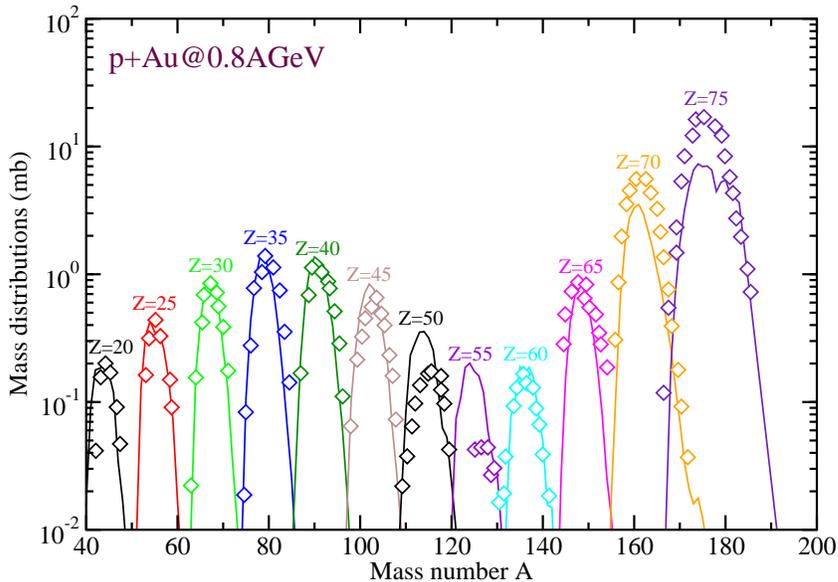}
\end{minipage}
\begin{minipage}[t]{16.5 cm}
\caption{Mass distributions for different isotopes in $p+Au@0.8~GeV$ reactions. 
Theoretical calculations (solid lines) are compared with experimental data 
(open diamonds) taken from \protect\cite{yieldsEXP}.
\label{Fig3}}
\end{minipage}
\end{center}
\end{figure}

Fig.~\ref{Fig2} shows double differential kinetic energy spectra of emitted 
neutrons as extracted from the combined GiBUU+SMM model and from the 
experiment. The comparison with the data is very good. In particular, the 
theoretical transport calculations reproduce all the details of the 
entire process, e.g., the quasi-elastic peaks at forward polar angles at 
high kinetic energies, which is a dynamical effect, and the low energy 
spectra, as the result of the statistical decay of the residual nucleus. 
Total fragment cross sections are reproduced reasonable within the combined 
model, in particular, for isotopes produced in the spallation region (near the 
target mass) and for fission fragments, see Fig.\ref{Fig3}.

\begin{figure}[tb]
\begin{center}
\begin{minipage}[t]{10 cm}
\epsfig{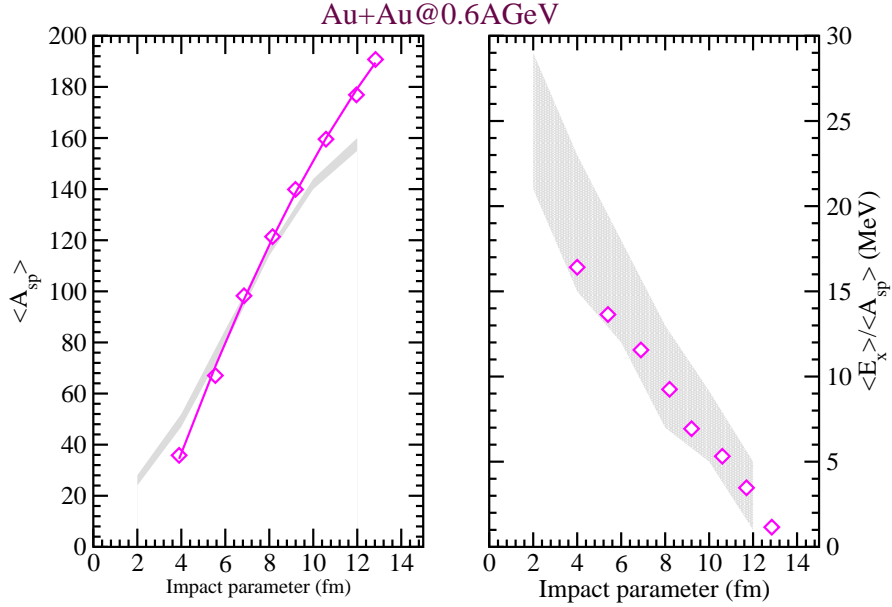}
\end{minipage}
\begin{minipage}[t]{16.5 cm}
\caption{Centrality dependence of the average mass number (left panel) and the 
excitation energy per nucleon (right panel) in projectile fragmentation in 
$Au+Au@0.6~AGeV$ reactions. Theoretical calculations (gray band) are compared 
with data from the ALADIN collaboration (open diamonds) \protect\cite{ALADIN}.
\label{Fig4}}
\end{minipage}
\end{center}
\end{figure}

In the more complex case of heavy-ion collisions the combined approach has been 
applied to spectator fragmentation only, since this system is well suited 
for theoretical and experimental studies of hypernuclei at higher energies (see 
below). As in the case of proton-induced reactions, the non-equilibrium dynamics 
of a heavy-ion collisions is modeled by the transport equation, until spectators 
achieve an equilibrated configuration characterized by a spherical local momentum 
distribution. Mass and charge number and excitation energy of spectators can be then 
determined, which serve as parameters for the statistical decay of excited 
spectators. 
Fig.~\ref{Fig4} shows the centrality dependence of the average mass and excitation 
energy per nucleon of spectator matter. The theoretical predictions for these 
quantities quantitatively fit the experimental data, which is an important issue 
when applying the statistical multifragmentation model. More successful comparisons 
with data can be found in Ref. \cite{gait08}.

We conclude that the combination between the dynamical non-equilibrium transport 
model and the statistical fragmentation approach is a reliable tool to study the 
fragmentation process in reactions. It is therefore natural to extend it by including 
strangeness degrees of freedom in the fragmentation process, which is the topic 
of the next section.

\section{Fragmentation at high energies (Hypernuclei)}

The production of hypernuclei in high energy reactions induced by protons, 
antiprotons and heavy-ions belong to the major projects in hypernuclear 
physics proposed by the HypHI- and PANDA-collaborations at the new FAIR 
facility at GSI \cite{hypHI,PANDA}. As a first testing phase the HypHI collaboration 
will start in the next year with heavy-ion experiments induced by 
light ${}^{12}C$ and ${}^{7}Li$ nuclei, and in the RIKEN facility with high 
energy proton beams on light ${}^{12}C$ targets. The reason of selecting 
collisions between light systems is the easier identification of 
hypernuclei via the weak decay of the hyperon into pions. 
In earlier theoretical studies cross sections of the order of only few 
microbarn ($\mu b$) were predicted \cite{wakai1}, due to the low cross sections 
of strangeness production and the rare effects of secondary scattering, important 
in producing slow hyperons inside the spectator regions. 

It is therefore natural to study these light systems again, in order to compare as 
a first step the results of our model with earlier theoretical and rare experimental 
predictions . The production of hypernuclei in spectator fragmentation has been 
theoretically modeled within a phenomenological coalescence prescription in 
coordinate and momentum space. The coalescence factors, which considerably influence 
the results, have been adjusted such to produce results as close as possible 
to earlier predictions of Refs. \cite{wakai1}. 
\begin{figure}[tb]
\begin{center}
\begin{minipage}[t]{10 cm}
\epsfig{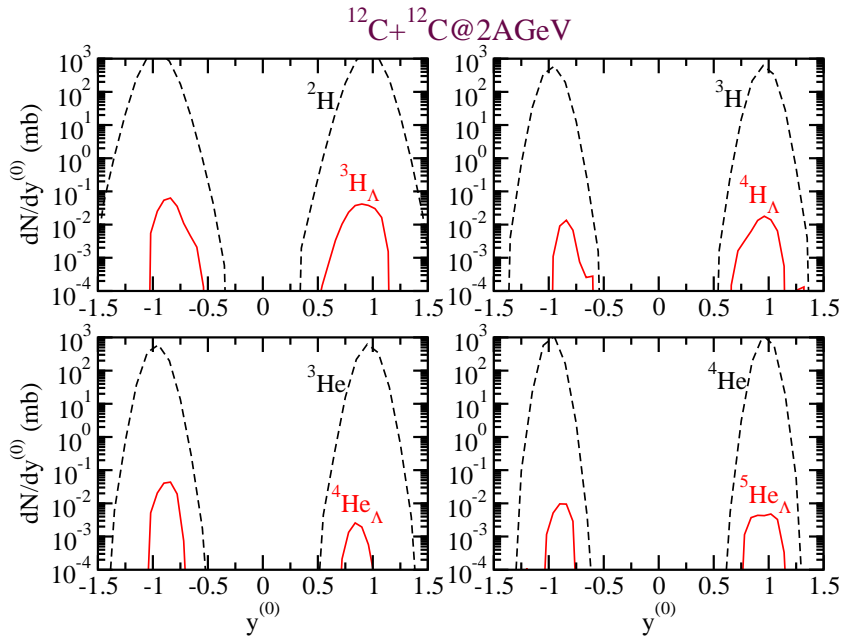}
\end{minipage}
\begin{minipage}[t]{16.5 cm}
\caption{Rapidity distributions as function of the rapidity 
$y^{(0)}$ (normalized to the projectile rapidity in the c.m. frame) of 
different particle types (as indicated) for the system 
${}^{12}C+{}^{12}C@2~AGeV$. 
\label{Fig5}}
\end{minipage}
\end{center}
\end{figure}
Fig.~\ref{Fig5} shows inclusive rapidity distributions of different light fragments 
and the corresponding hyperfragments in spectator fragmentation. The 
estimated hyperfragment production is ca. $5-6$ orders of magnitude less than 
that of fragment production in general. This effect was expected due to two 
reasons: (a) the low values of strangeness production cross sections with $3,4$-body 
final states $BB\rightarrow BYK$, 
$BB\rightarrow BBK\overline{K},~B\overline{K}\rightarrow YB$ \cite{tsushima}, 
which are important in producing {\it slow} hyperons, 
(b) and the small interaction spectator volume which prevents hyperon production inside the 
spectator matter in secondary processes, e.g., $B\pi\rightarrow YK$ and 
$B\overline{K}\rightarrow \pi Y$, and their elastic re-scattering. An 
integration over the projectile rapidity leads to total production cross 
sections of light hypernuclei $\sigma_{tot}=2.2,~4,~1.4~\mu b$ for 
${}^{4}_{\Lambda}H,~{}^{4}_{\Lambda}He$ and ${}^{5}_{\Lambda}H$ single 
hypernuclei, respectively, in spectator fragmentation. In particular, the 
contribution to hyperfragment formation from secondary pion-nucleon scattering 
inside the spectators is very moderate, e.g., $\sigma_{\pi N}=0.3,~0.2,~0.03~\mu b$ 
for the formation of ${}^{4}_{\Lambda}H,~{}^{4}_{\Lambda}He$ and ${}^{5}_{\Lambda}H$, 
respectively. These results, which are consistent with earlier studies from 
Wakai et al. \cite{wakai1}, lead to the conclusion that the major contribution to 
hypernuclear production originates from the capture of fireball hyperons during 
the passage stage of the spectators near the expanding fireball region. 

\begin{figure}[tb]
\begin{center}
\begin{minipage}[t]{10 cm}
\epsfig{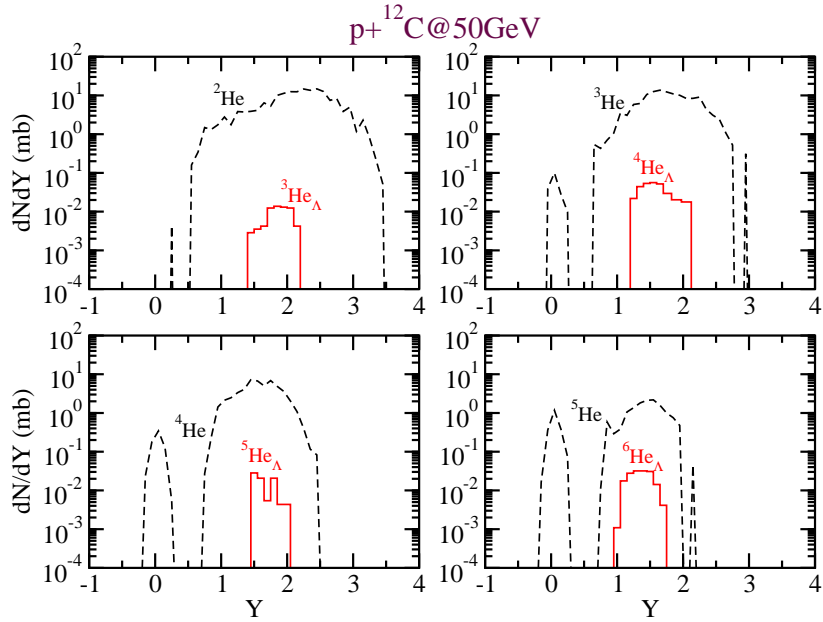}
\end{minipage}
\begin{minipage}[t]{16.5 cm}
\caption{Rapidity distributions of different particle types (as indicated) 
for the system $p+{}^{12}C@50GeV$. 
\label{Fig6}}
\end{minipage}
\end{center}
\end{figure}

In proton-induced reactions at much higher energies, e.g., $p+C@50~GeV$ 
(J-PARC) \cite{JPARC}, the situation turns out to be different, as found in 
dynamical transport calculations. The major channels contributing to the 
formation of light hypernuclei are those with pion-baryon scattering. This 
seems obvious due to the very high total pion production cross section, 
in contrary to the $C+C@2~AGeV$ colliding system. The situation is summarized 
in Fig.~\ref{Fig6}, again in terms of the rapidity distribution. An interesting 
feature here is the appereance of two sources, a target at rest and a moving 
source. This dynamical break-up is due to small re-scattering effects inside 
the initial compound nucleus. A significant amount of the beam energy is thus 
transferred into only a few nucleons, which causes the pre-equilibrium break-up. 
The transport theoretical results predict high energetic light hypernuclei, which 
might be easily separated from the background and thus be experimentally 
accessible. 

\section{Final remarks}

The investigation of hypernuclei in reactions, as they will be experimentally 
studied in the new FAIR facility at GSI, provides new insight on the 
still controversial hyperon-nucleon and hyperon-hyperon interaction. It 
has been thus obvious to explore this field from the theoretical point 
of view, as has been done here in terms of covariant transport dynamics. 

We have theoretically explored the different mechanisms for hypernuclear 
production in heavy-ion and proton-induced reactions relevant to the 
future experiments at the new FAIR facility at GSI on hypernuclear 
physics, which have to be compared with experimental data, when they will 
be available. 

Of particular interest will be the study of double-strange hypernuclei or 
generally from exotic multi-strange bound objects, e.g., 
${}^{A}X_{\Lambda\Lambda},~{}^{A}X_{\Omega}$, which is an important 
issue in theoretical and experimental works to better understand the 
hyperon-hyperon force. This project, which will be one part of the proposals of 
the PANDA-collaboration, will be theoretically studied in antiproton-induced 
reactions. We conclude that the present theoretical work gives an appropriate 
basis for investigations on hypernuclear physics at the FAIR facility at GSI.


\section*{Acknowledgement}
\itemsep -2pt 
We would like to thank for many useful discussions 
A. Botvina. A.B. Larionov and I.N. Mishustin. We thank T. Saito and also 
the members of the HypHI Collaboration for their proposal to study the 
colliding systems. Finally, we would like to thank the GiBUU group for 
useful discussions. This work is supported by BMBF.



\begin{thebibliography}{99}
\bibitem{residual}
The residual nucleus is defined as a compound system of particles 
with densities greater than $\rho_{cut}/\rho_{sat}$ with 
$\rho_{cut}=100$ and $\rho_{sat}=0.16~fm^{-3}$. Thus, the parameters 
of a residual nucleus (or fragmenting source) are given by the 
mass ($A$) and charge ($Z$) numbers and the excitation energy.
\end{thebibliography}
\end{document}